# DESIGN OF THE EURISOL MULTI-MW TARGET ASSEMBLY: RADIATION AND SAFETY ISSUES

**Marta Felcini** [1] (a,d), **Adonai Herrera-Martínez** (a),

**Yacine Kadi** (a), **Thomas Otto** (b) and **Luigi Tecchio** (c)

(a) CERN Accelerators and Beams Department, Geneva, Switzerland
(b) CERN Safety Commission, Geneva, Switzerland
(c) INFN Laboratori Nazionali di Legnaro, Legnaro Italy
(d) University of California Los Angeles, Physics Department, Los Angeles USA

[1] Presenting author

## Abstract

The multi-MW target proposed for the EURISOL facility will be based on fission of uranium (or thorium) compounds to produce rare isotopes far from stability. A two-step process is used for the isotope production. First, neutrons are generated in a liquid mercury target, irradiated by the 1 GeV proton or deuteron beam, provided by the EURISOL linac driver. Then, the neutrons induce fission in a surrounding assembly of uranium carbide. R&D projects on several aspects of the target assembly are ongoing. Key criteria for the target design are a maximum beam power capability of 4 MW, a remote handling system with minimum downtime and maximum reliability, as well as radiation safety, minimization of hazards and the classification of the facility. In the framework of the ongoing radiation characterization and safety studies, radiation transport simulations have been performed to calculate the prompt radiation dose in the target and surrounding materials, as well as to determine shielding material and angle-dependent parameters. In this paper, we report the results of these studies and the proposed radiation shield design for the multi-MW target area. Furthermore, accurate estimates have been performed of the amount of fissile elements being produced in the uranium target assembly, for typical running conditions, in order to understand the implications for the classification of the facility. The results are reported and briefly discussed.

Invited talk at the SATIF-8 Workshop, Pohang Accelerator Laboratory, May 22-24, 2006.

Project supported by EC under EURISOL DS Contract no. 515768 RIDS, www.eurisol.org .

**Introduction**

The proposed European Isotope Separation On-Line (EURISOL) facility [1] is an ISOL-type next-generation Radioactive Ion Beam (RIB) facility which will provide RIB intensities two to three orders of magnitudes higher than in existing RIB facilities. The EURISOL physics programme will address fundamental questions related to nuclear structure and many-body interactions between hadrons, as well as the precision study of nuclear interactions, which determine element formation in stars and have important implications in the understanding of stellar and galaxy evolution. Furthermore, the availability of such a high-intensity radioactive ion source will open up new opportunities to test fundamental symmetries of the Standard Model at low energy, which are complementary to those performed with high-energy experiments. The possibility of producing a pure beam of electron neutrinos (or their antiparticles) through the beta-decay (beta beam) of radioactive ions circulating in a high-energy storage ring, for neutrino precision experiments, is also under study.

In the ISOL method, isotopes are produced by spallation or fission reactions in thick targets. This method is complementary to in-flight isotope production, used for instance in the FAIR project at GSI (Darmstadt, Germany). In the EURISOL Design Study (DS) [2], two baseline designs for isotope production targets are considered: a direct target and a fission target.

The direct target is the extension to higher beam power of existing ISOL targets at ISOLDE (CERN, Switzerland), ISAC (TRIUMF, Canada) or SPIRAL (GANIL, France). A proton beam deposits a power of up to 100 kW in a EURISOL direct target. There is a physical limitation to the maximum power rating of such a target, which is given by the capability to shed excess heat and to maintain a temperature in the target below the target limit for thermal disintegration.

A fission target coupled to a spallation neutron source is a way to overcome this power limit of the direct targets. This is the option considered for the multi-MW target assembly of the EURISOL facility. A multi-MW 1 GeV proton beam generates spallation neutrons in a liquid mercury target. The neutrons in turn induce fission reactions in an actinide target assembly surrounding the spallation neutron source. The liquid mercury target can withstand and evacuate much higher beam power than a solid direct target. The isotope production efficiency (expressed as isotopes per unit beam power) of the fission target depends largely on the geometry of the spallation neutron source, the actinide target and the neutron-moderating or reflecting materials Similar efficiencies as for direct targets can be achieved. The target material for the fission target is a low-density actinide-carbide, for example $UC_x$ or $ThC_x$, where *x* lies between 3 and 4.

In contrast to the direct actinide target, where the dominating reactions are spallation and fast fission with *protons*, isotope production in a fission target is dominated by *neutron* reactions: fission of $^{238}U$ with fast neutrons, and of $^{235}U$ with epithermal and thermal neutrons. A third neutron-induced reaction, neutron capture on $^{238}U$, leads to the breeding of fissile elements in the target, the most important being $^{239}Pu$. The multi-MW target assembly is described in Sect. 2.

R&D projects on several aspect of the target assembly are ongoing. Key criteria for the target design are a maximum beam power capability of 4 MW, remote handling system with minimum downtime and maximum reliability, as well as radiation safety, minimization of hazards and the classification of the facility.

In the framework of the ongoing radiation characterization and safety studies, detailed radiation transport simulations have been performed to calculate the prompt radiation dose in the target and surrounding materials, as well as determine shielding material and angle-dependent parameters. We report in Sect. 3 the results of these studies and the proposed radiation shield design for the multi-MW target area. Furthermore, estimates have been performed of the amount of fissile elements being produced in the uranium target assembly for typical running conditions. in order to understand the implications for the classification of the facility. These results are reported in Sect. 4 and briefly discussed. Sect. 5 presents a summary of the results reported in this paper

**The multi-MW target assembly**

*Geometry*

The multi-MW target assembly consists of a spallation neutron source, converting the proton beam from the accelerator into neutrons, and a target in which a fissile isotope undergoes reactions with fast or with thermal neutrons, producing radioactive isotopes as fission products.
The present target design [3,4] is preliminary. However it is detailed enough to make meaningful radiation characterization studies and predictions on the production rates within the target of radioactive isotopes, including actinides.

The spallation neutron source is a 50 cm long, 16 cm diameter cylindrical steel vessel filled with mercury (Hg) [3]. Two ring-shaped tantalum containers with tungsten heat screens are placed around the cylindrical neutron source. The targets are filled with pellets made from uranium carbide (stochiometric notation: $UC_x$, with $x \approx 3$) with a density of 3 g cm$^{-1}$, occupying 85 % of the volume [4]. Tab.1 shows the dimensions and mass of the $UC_x$ targets. Fig.1 is an isometric view of the target and Fig. 2 an assembly drawing. The $UC_x$ target containers are held in place with supports made from beryllium oxide (BeO). A 10 cm thick graphite moderator encloses the spallation neutron source plus fission target assembly, azimuthally and downstream, perpendicularly to the beam axis

**Table 1 Dimensions and $UC_x$ filling parameters of the two ring-shaped fission targets [4] placed around the spallation neutron source.**

|  | small radius target | large radius target | total |
|---|---|---|---|
| Length | 20 cm | | |
| Thickness | 2.8 cm | | |
| inner radius | 12.1 cm | 19.1 cm | |
| Volume | 4750 cm$^3$ | 7213 cm$^3$ | 11963 cm$^3$ |
| $UC_x$ density, filling factor | 3.0 g cm$^{-3}$, 85 % | | |
| $UC_x$ mass | 12.1 kg | 18.4 kg | 30.5 kg |

*Target material*

The actinide target material to be used in the multi-MW fission target has not been decided yet. In present direct isotope production targets, depleted uranium is the most frequently used actinide. This material accounts for 50% of all targets used at ISOLDE (CERN). Depleted uranium is a by-product of nuclear fuel production. It contains 99.7% $^{238}$U, 0.3% $^{235}$U and a smaller amount of $^{234}$U. It is well suited for the production of direct targets where the fission reactions are caused mainly by fast protons and secondary particles (neutrons, pions). For the EURISOL multi-MW fission target assembly, the choice of natural uranium (0.7% of $^{235}$U) as target actinide would have advantages, because it would better exploit the epithermal and thermal neutrons emerging from the spallation source and from the moderator.

In a realistic setting, the fission target assembly will be embedded in massive radiation shielding. The shielding must be considered when estimating isotope production and fissile isotope breeding rates because it moderates and reflects neutrons from the spallation source, increasing the efficiency of isotope production per primary beam proton. In this paper, a cylindrical steel shielding around the target assembly has been considered, as described in the next Section. Upstream of the target, the shielding is closed by an end-cap, leaving only a narrow penetration for the proton beam line. Downstream of the target, a mobile shielding end-cap has to be arranged, allowing retrieving the target assembly to a (shielded and confined) maintenance position.

**Figure 1: Isometric view of the fission target assembly. A mercury target, serving as spallation neutron source (green), is surrounded by two ring-shaped target containers filled with uranium carbide, as fissile material (grey). Isotopes created in the fission targets move by diffusion through evacuated tubes (blue) to the ion source. A graphite moderator (grey) is enclosing the spallation neutron source plus fission target assembly. The whole target is mounted in a stainless steel (SS) vacuum chamber (light grey).**

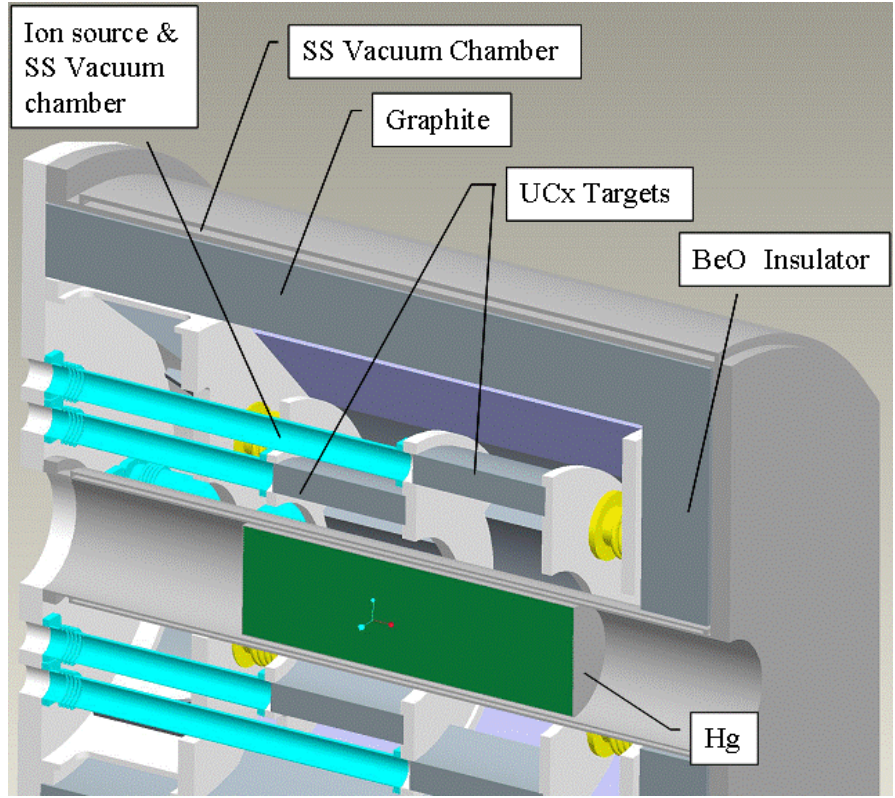

**Figure 2: Assembly drawing of the fission target assembly. Length and diameter of the assembly are approximately 100 cm each.**

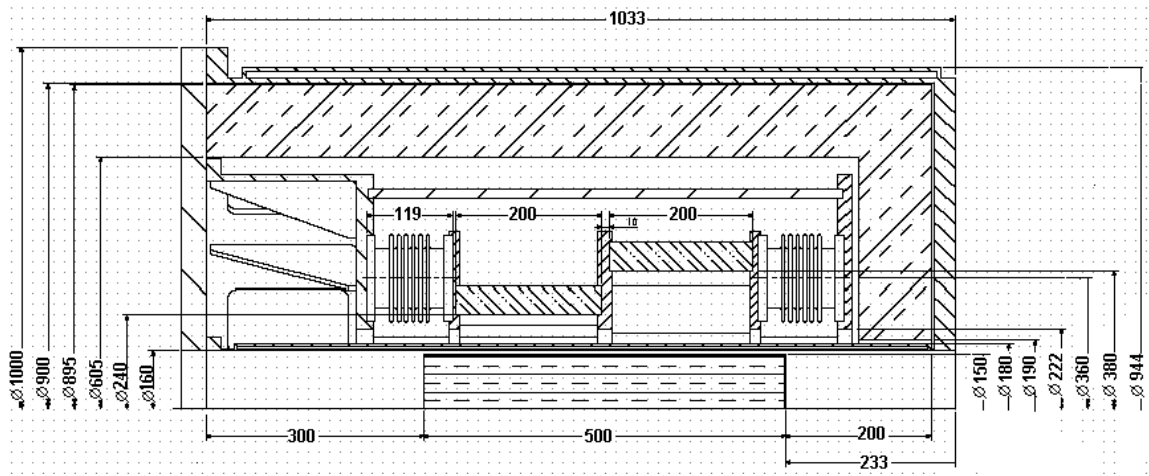

*Simulation*

The Monte Carlo (MC) radiation transport program FLUKA [5] is used to perform simulations of the dose rate fields in the multi-MW fission target assembly and the surrounding materials, as well as for the calculation of isotope production. The main components of the fission target assembly and the shielding are described with the combinatorial geometry package of FLUKA. The isotope production rates are determined by the fluence spectra of secondary particles (neutrons, pions, protons) from the spallation source in the fission target material.

In the MC simulation, a Gaussian-profiled proton beam with FWHM = 3.54 cm impinges on the centre-line of the mercury target. The production of isotopes ("residual nuclei") is scored with the RESNUCLE card, employing models for interactions of nuclei with hadrons and tabulated cross sections for interactions with neutrons ok kinetic energy $E_{kin} < 19.6$ MeV.

**Dose rates and shielding design for the multi-MW target station**

One of the technological challenges of this multi-MW facility is the very intense radiation field created by the secondary particles produced in the target and the surrounding materials. This imposes stringent constraints on the construction, as well as on the operation and maintenance of the target and its associated equipment. To ensure safe operation, the target station must be encapsulated in a thick shield to attenuate the radiation dose outside the shield to acceptable values. The shield around the target must be designed to satisfy a number of requirements. The shield must be sufficiently thick, but at the same time the amount of material used for the shield must be minimised, to reduce construction and disposal costs. The volume inside the shield around the target must also be minimised to facilitate evacuation of the target area and avoid air activation. At the same time, the small distance between the target and the shielding material implies that (a) the radiation level at the target area is increased by the back-scattered particles, (b) high power density is deposited in the inner shield layer, requiring shield cooling and (c) the inner shield layer will be highly activated. With these constraints in mind, the shielding materials, thickness and shape must be optimally chosen.

The purpose of the prompt radiation shield is to reduce the radiation dose outside the shield to values below the operational dose rate limits defined by the legislation. They are different in different countries. Since we do not know in which country the EURISOL facility will be sited, we have chosen the limit of 1 μSv / h, as a typically permissible dose rate in supervised radiation areas. Thus, our objective is to design a prompt radiation shield for the multi-MW fission target, in the configuration as described in the previous Section, exposed to 1 GeV 4 MW proton beam, such that the dose rate at the outer surface of the shield is inferior to 1 μSv/h.

For the purpose of shield design, given the beam power and energy, the effective dose rate **H** from prompt radiation can be approximated by the point source, line-of-sight attenuation formula

$$\mathbf{H}(\theta, d, R) = \mathbf{H}_0(\theta) \exp\{-\Sigma_i \, d_i \, / \, \lambda_i \, (\theta)\}/R^2 \,, \tag{1}$$

where $\theta$ and **R** are the coordinates (azimuthal angle and distance) of the point where the dose rate is measured and **d**=$\Sigma_i$ **d**$_i$ is the total distance traversed by the radiation in different successive shielding materials. **H$_0$** ($\theta$) is the source term, depending upon the azimuthal angle $\theta$. It gives the dose rate at a unit distance from the source before any shielding material and it depends on the source intensity and geometry. Once the source term and the attenuation lengths $\lambda_i$ in the different shielding materials are known, it is possible to determine, as a function of $\theta$, the shield thickness **d$_i$** of the different materials, needed to attenuate the radiation dose by the desired factor. It is useful to remind that the attenuation length in a given material depends upon the composition in terms of particle types (protons, neutrons, pions, etc) and energies of the radiation field.

Our strategy [6] to design the radiation shield for the multi-MW target is the following. We use the FLUKA MC code to simulate the particle transport through the target and shielding materials and calculate the dose rate spatial distribution H(θ,R), in a reference shield geometry. Then we fit the parameters **$H_0$ (θ)** and **λ $_i$(θ)** for subsequent determination of the angular-dependent extension **$t_i$(θ)** of the shielding shells, making use of eq.(1).

The shield reference configuration is a two-component layered steel and concrete cylindrical monolith, completely encapsulating the target. It is composed of an inner steel cylinder of 2 m constant thickness, of 60 cm inner radius, closed at the two ends by 2 m thick end-caps, sited at about 20 cm distance from the target ends. The steel layer is followed by a 2 m thick concrete layer, also cylindrical in shape, closed by 2 m thick end-caps. The chosen thickness values are sufficiently large to allow establishing particle equilibrium, after which the radiation field composition, in terms of particle types, stays unchanged with the penetrated material depth. Consequently, a reliable determination of the source term and attenuation length values can be performed.

The iso-dose contours, as a function of the longitudinal coordinate z and the radial coordinate r (cylindrical symmetry is assumed in the simulation) are shown in the upper panel of Fig.3. Superimposed to the dose contours are the target and shield geometries as used in the simulation. The subdivision of the steel shield in 20 layers and of the concrete shield in 10 layers indicates the regions where suitable MC biasing factors have been applied, to achieve adequate particle statistics at all depths in the shielding materials. After conversion from cylindrical **(r,z)** to spherical coordinates (**$R=(r^2+z^2)^{1/2}$, θ = arccos(z/R)**) dose rates as a function of **R** for two different values of **θ,** in the beam direction (θ = 0 degrees) and perpendicular to the beam direction (θ = 90 degrees) are shown in the lower panels of Fig.3. In the two cases, the contributions to the dose rate from the different particle types are also shown.

The values of dose rate times **$R^2$**, **$HR^2$**, are then plotted as a function of the radiation depth **d** in the shielding material (**d (θ)=R-g(θ)**, **g(θ)** being the distance of the shield inner edge from the coordinate origin), in different θ bins of 4 degree widths, going from 0 (beam direction) to 180 degrees (beam opposite direction). Then, these curves are fitted in each material using an exponential function **exp(a+bd)** and the values of the source term and attenuation length derived from the fitted values (**$H_0$=exp(a)**, λ=**-1/b**). As an example the attenuation curves in steel for θ values between 0 and 90 degrees are shown in the upper panel of Fig.4. The values of **$H_0$** and λ resulting from the fit in steel as a function of θ are given in the lower panel of Fig.4. The larger values of the source term **$H_0$** in the backward direction (θ > 90 degrees) may seem unnatural, as one would expect lower dose values in the backward direction. This is indeed the case, as lower λ values in the backward direction compensate the higher **$H_0$** values. In the fit, we have chosen deliberately to minimize the uncertainties on the λ parameters, while allowing relatively bigger uncertainties on the **$H_0$** parameters [6]. In this way, as λ enters exponentially, while **$H_0$** enters linearly in eq. (1), we minimise the uncertainty on the calculated **H(θ,d,R)** values induced by the uncertainties on the **$H_0$** and λ fitted values. Thus, the retained **$H_0$** and λ values, shown on the bottom of Fig.4 for steel, are those which give the best fit, while minimising the uncertainties on the λ parameters.

The λ values have also been determined in the concrete shield and used, together with the fitted **$H_0$** and λ parameter values in steel, to evaluate the thickness of steel and concrete, as a function of **θ,** such that the dose rate at the outer surface of the shield is inferior to 1 μSv/h. The result is shown in Fig. 5. The needed concrete thickness **$t_{conc}$** is given as a function of θ for different values of the steel thickness **$t_{steel}$**. Here **$t_{steel}$** is defined as the thickness of the steel cylinder, along the axis and of the end caps, encapsulating the target. The inner dimensions of the steel cylinder are the same as described above, for the reference shield geometry.

**Figure 3: (Top) Iso-dose rate curves as a function of r and z (cylindrical coordinates used for the simulation). Effective dose rate values, in Sv/h, are given here and in the following Figures for 4 MW 1 GeV proton beam impinging on the target. (Bottom) Effective dose rate as a function of $R = (r^2+z^2)^{1/2}$ for two different values of θ, in the beam direction (θ = 0 degrees) and perpendicular to the beam direction (θ = 90 degrees). In the two cases, the contributions to the dose rate from the different particle types are also shown.**

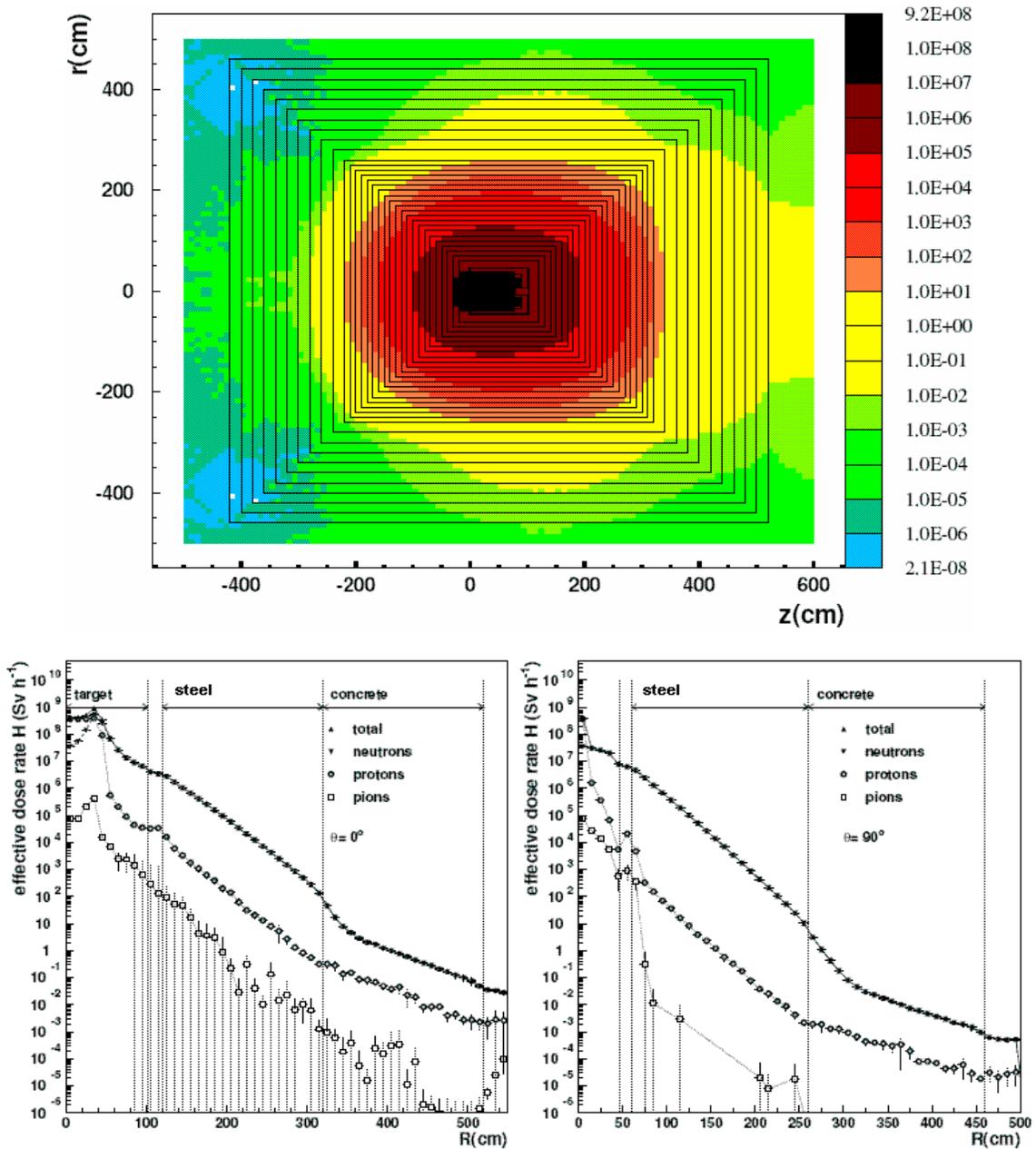

**Figure 4: (Top) Curves of effective dose rates times $R^2$ in steel, as a function of steel depth, $d_{steel}$, for different θ values. (Bottom) The values of $H_0$ and λ, obtained from the fit of the attenuation curves, are shown as a function of θ.**

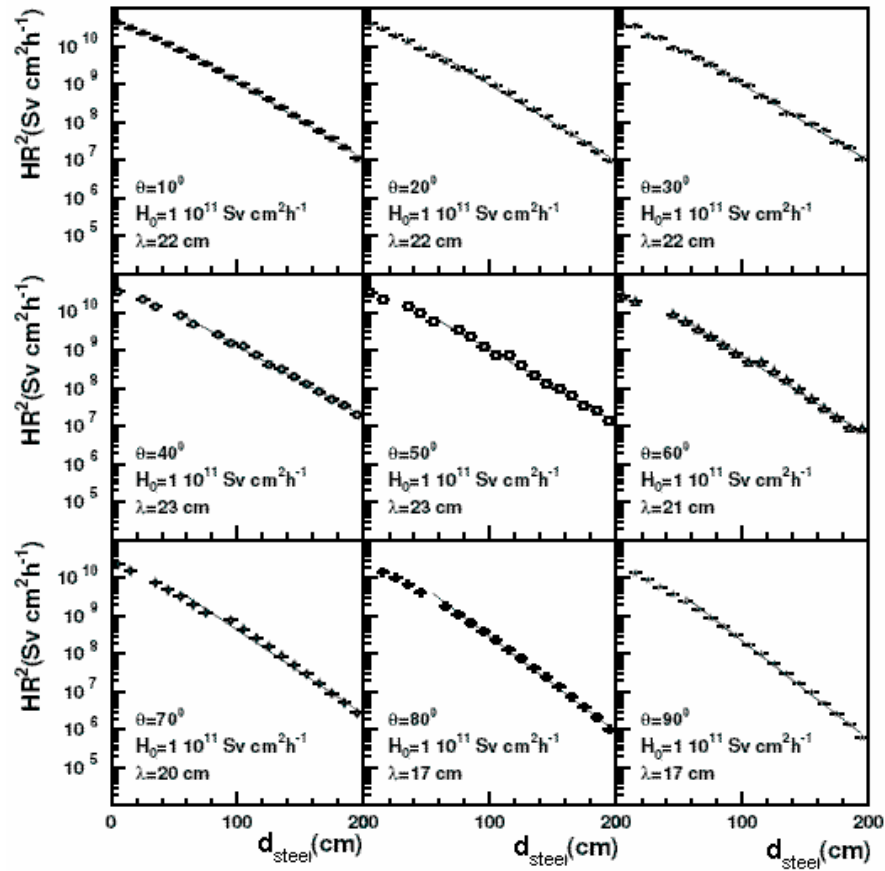

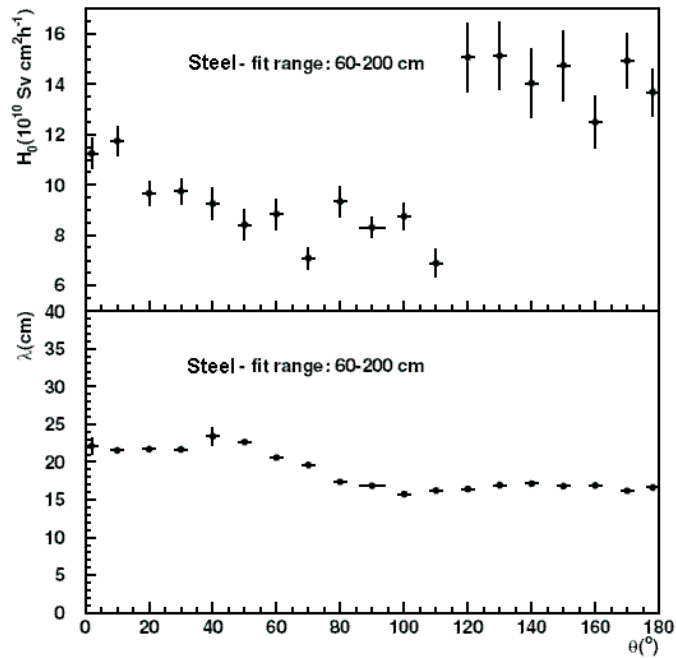

**Figure 5:** Concrete shield thickness, as a function of the azimuthal angle θ, needed to attenuate the effective dose rate to 1 μSv/h at the outer shield surface. The curves are given for six different values of the inner steel cylinder thickness $t_{steel}$ (see text). The upper curve is for $t_{steel}$ = 100 cm while the lower curve is for $t_{steel}$ = 300 cm, with the intermediate curves given for intermediate $t_{steel}$ values in the order shown in the legend. The points are the result of the dose rate calculation, eq. (1), using the source term and attenuation length values, obtained from the fit of the simulation data, as explained in the text. The curves are the results of a polynomial fit to the points and their uncertainties

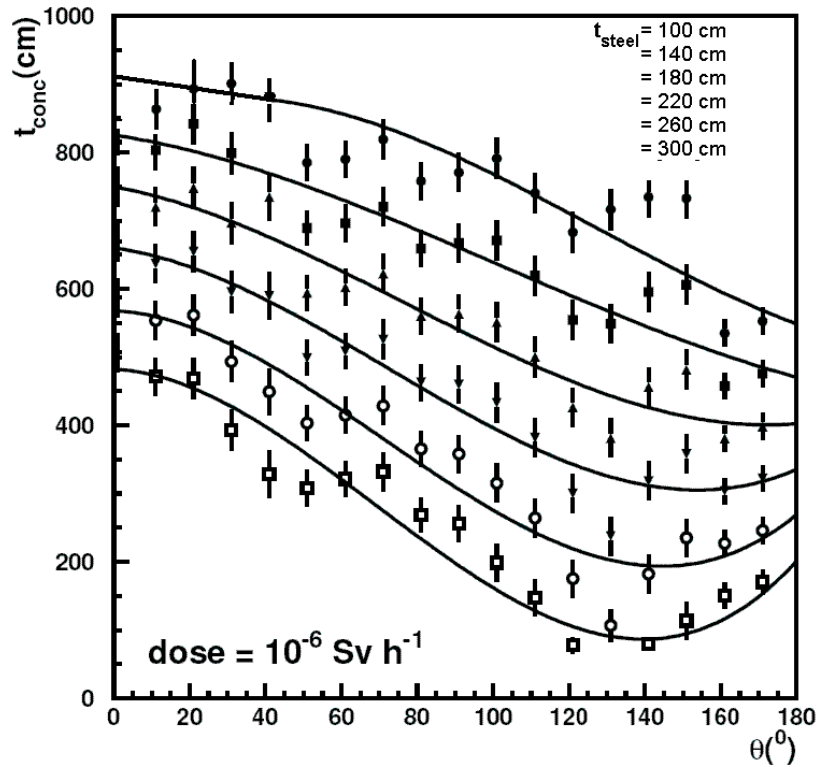

Given the cost of steel, currently about a factor of 25 more expensive than concrete, we propose to use 1 m thick steel cylinder followed by a concrete shield of variable thickness depending on θ, as indicated in Fig.5 for $t_{steel}$ = 1m (upper curve). For θ = 0, 90 and 180 degrees a concrete thickness values of about 900, 800 and 550 cm, respectively, are needed after 1 m thick steel shield cylinder, to reduce the dose rate to 1 μSv / h.

**Breeding of fissile material in the multi-MW target**

Fissile isotope breeding in five different fission target designs has been studied [7]. Their characteristics are described in the second column of Tab. 2. The standard ISOLDE $UC_x$ target has been added as an example for direct targets. The designs U0 and Th0, in absence of shielding, although unrealistic, allow drawing conclusions on the relative merits of different actinide materials for isotope production and breeding of fissile elements. Comparison between the designs U0 and U1 will reveal the influence of the shielding. The designs U1, U2 and U3 can be considered as realistic, keeping in mind that the different $^{235}$U-content will have effects on isotope production (as shown in the fifth column of Tab.2).

The method used for the calculation is described in detail in [7]. The results in terms of estimates of fissile isotope production are reported in the last two columns of Tab. 2

The studied fission target assemblies show approximately the same efficiency (see column 5 of Tab.2) as a standard ISOLDE $UC_x$ target in the production, per incident proton, of spallation and fission elements lighter than the target material. Reducing the size of the fission target would reduce this efficiency. As it can be seen in Tab.2, one isotope alone, $^{239}U$, is produced more copiously than all others taken together, in the case of an $UC_x$ target. Essentially all of it will decay into $^{239}Pu$. In direct targets (the ISOLDE target taken as an example), breeding of heavier isotopes is practically absent because only few low-energy neutrons are available for its production. The presence of the shielding, in the configuration described in the previous Sections, increases the production of $^{239}U$ (and thus $^{239}Pu$) by 70 %. This figure may vary depending on the exact arrangement of shielding. The higher $^{235}U$-content enhances the production of fission products due to a better exploitation of epithermal and thermal neutrons, but it does not influence the neutron spectrum and capture rate, thus the breeding of $^{239}U$ and $^{239}Pu$.

In the Th-based fission target, slightly more fissile products are bred than in the U-based target, for a lesser amount of light (A<239) isotopes produced per incident proton. The amount of $^{233}U$ produced in a Th-based target presents similar risks for radiation protection and same issues for nuclear security as the $^{239}Pu$ produced in a U-based target. In this respect there is no advantage in considering Th as base material for fission targets. The comparison with a standard ISOLDE target, also shows that the production of fissile elements, as compared to that of lighter isotopes, is highly reduced in the direct target with respect to the fission target assembly.

In summary, in multi-MW target assemblies for isotope production based on uranium carbide, the isotope $^{239}U$ is produced by neutron capture on $^{238}U$ more abundantly than all other isotopes taken together. Practically all of the $^{239}U$ will decay into $^{239}Pu$, the burn-up of this isotope in the target being very low. Similar conclusions apply to targets based on thorium carbide for the production of the fissile isotope $^{233}U$. In a realistic target design, including the effects of radiation shielding on the neutron fluence, more than 70 g of $^{239}Pu$ would be produced per year, irrespective of the presence of $^{235}U$ in the target material (see Tab.2). Rules for safeguarding fissile materials under the international nuclear Non-Proliferation Treaty (NPT) will have to be applied to the EURISOL multi-MW targets.

**Conclusions**

The EURISOL facility will provide RIBs with intensities two to three orders of magnitudes higher than in present facilities, opening up unprecedented opportunities for investigations and discoveries in nuclear physics, astrophysics and particle physics.

This paper focus on the EURISOL multi-MW target, designed for the production of rare isotopes far from stability. The proposed design makes use of a liquid mercury target, directly irradiated by a multi-MW proton beam, for the production of an intense neutron flux. The neutrons in turn irradiate a surrounding assembly of uranium carbide targets in which isotopes are produced by neutron-induced fission reactions. A detailed layout of the multi-MW target assembly has been proposed and its key features are being studied.

Prompt radiation dose calculations in the target and the surrounding materials have been performed, the radiation attenuation through materials has been analysed and its angular dependence parameterised. It has been shown that with a 1 m thick steel monolith encapsulating the target, followed by 5.5 m (up-stream of the target) to 9 m (down-stream of the target) thick concrete shield, the radiation dose outside the shield can be kept below the limit of 1 μSv/h, as permissible dose rate in supervised radiation areas.

In terms of nuclear safety, the quantities of hazardous isotopes have been calculated, showing no particular advantage in using Th as fissile material rather than U, due to the breeding of $^{233}U$, as problematic for proliferation as $^{239}Pu$. Moreover, Th-based targets present lower production per proton of useful isotopes, than U-based targets. These results are an essential input in the finalisation of the target design and in the process leading to the classification of the facility.

Table 2: Estimate of fissile isotope production in a $UC_3$ ($ThC_3$) target, for five different configurations [7]. Columns 2, 3 and 4 show the characteristics and shielding conditions used for the calculation (see text). Column 5 gives the production per proton of isotopes lighter than $^{239}U$. Columns 6 and 7 give the production of $^{239}U$ ($^{233}Th$ for Type Th0) per proton and the implied production of $^{239}Pu$ ($^{233}U$ for Type Th0) during 3000 hours with a 4 MW proton beam. Isotope production in an ISOLDE-$UC_x$ target is added for comparison.

| Target Type | Common characteristics | Actinide used for target | Steel Shield | Light (A<239) isotopes per proton | $^{239}U^{(*)}$ per proton | $^{239}Pu^{(*)}$ for 4 MW power 3000 h running time | |
|---|---|---|---|---|---|---|---|
| | | | | atoms | atoms | atoms | mass(g) |
| U0 | Target material $UC_3$ ($ThC_3$), Target mass 30.5 kg, 10 cm C moderator | $^{238}U$ | no | 0.31 | 0.407 | $1.09\ 10^{23}$ | 43.4 |
| Th0 | | $^{232}Th$ | no | 0.21 | 0.427 | $1.15\ 10^{23}$ | 45.6 |
| U1 | | $^{238}U$ | yes | 0.32 | 0.68 | $1.83\ 10^{23}$ | 72.5 |
| U2 | | $U_{nat}$ (0.7% $^{235}U$) | yes | 0.47 | | | |
| U3 | | $U_{dep}$ (0.3% $^{235}U$) | yes | 0.44 | | | |
| ISOLDE | Target material $UC_3$, dir. irrad. by $E$=1.4 GeV proton beam | $U_{dep}$ (0.3% $^{235}U$) | no | 0.33 | $9.6\ 10^{-4}$ | - | - |

$(*)$ $^{233}Th/^{233}U$ for Type Th0

**References**

[1] The EURISOL Collaboration, "The EURISOL Report – A feasibility study for a European Isotope Separation On-line Radioactive Ion Beam Facility ", European Commission Contract No. HPRI-CT-1999-50001, Dec. 2003, available at http://www.ganil.fr/eurisol/Final_Report.html

[2] EURISOL Design Study, http://www.eurisol.org/site01/index.php.

[3] A. Herrera-Martinez and Y. Kadi, "EURISOL-DS Multi-MW Target Preliminary Study of the Liquid Metal Proton-to-Neutron Converter", CERN-AB-Note-2006-013, 2006, and "Neutronic Calculations for the Baseline Configuration of the Multi-MW Mercury Target", EURISOL DS/TASK2/TN-05-03,2005.

[4] L. Tecchio et al., EURISOL DS WP4 Note in preparation.

[5] A. Fasso`, A. Ferrari, J. Ranft, and P.R. Sala, "FLUKA: a multi-particle transport code", CERN-2005-10 (2005), INFN/TC_05/11, SLAC-R-773 FLUKA , and references therein.

[6] M. Felcini, "Optimised radiation shield design for the EURISOL multi-MW target station", CERN-AB-Note-2006-015, 2006.

[7] Th. Otto, "Breeding of Fissile Isotopes in EURISOL Multi-Megawatt Targets", Report CERN-SC-2006-020-RP-IR., 2006.